\documentclass[12pt,a4paper]{article}
\pdfoutput=1
\usepackage{jheppub}
\usepackage{amsmath,amssymb,euscript,array,mathrsfs,epsfig}
\usepackage[small]{caption}

\def\ben
{\begin{equation}}
\def\een{\end{equation}}

\let\a=\alpha   \let\d=\delta 
  \let\q=\theta \let\k=\kappa
\let\l=\lambda \let\m=\mu \let\n=\nu

    \let\L=\Lambda
 \let\P=\Phi  
 \let\W=\mu

\let\fr=\frac
\def\W={\cal W}
\def\L ={\cal L}

\let\pa=\partial
\def\be{\begin{equation}}
\def\ee{\end{equation}}
\def\ba{\begin{array}}
\def\ea{\end{array}}

\def\dalemb#1#2{{\vbox{\hrule height .#2pt
        \hbox{\vrule width.#2pt height#1pt \kern#1pt
                \vrule width.#2pt}
        \hrule height.#2pt}}}

\newcommand{\bea}{\begin{eqnarray}}
\newcommand{\eea}{\end{eqnarray}}

\title{Holographic flows and thermodynamics of Polyakov loop impurities}
\author{S. Prem Kumar}
\author{and Dorian Silvani}
\affiliation{Department of Physics,\\Swansea University, \\
Singleton Park,\\ Swansea, SA2 8PP, U.K.}
\emailAdd{s.p.kumar@swansea.ac.uk, d.silvani.492808@swansea.ac.uk}
\abstract{We study holographic probes dual to heavy quark impurities interpolating between fundamental and symmetric/antisymmetric tensor representations in  strongly coupled ${\cal N}=4$ supersymmetric gauge theory. These correspond to non-conformal D3- and D5-brane probe embeddings in AdS$_5\times {\rm S}^5$ exhibiting flows on their world-volumes. By examining the asymptotic regimes of the embeddings and the one-point function of static fields sourced by the boundary impurity, we conclude that the D5-brane embedding describes the screening of fundamental quarks in the UV into an antisymmetric source in the IR, whilst the non-conformal, D3-brane solution interpolates between the symmetric representation in the UV and fundamental sources in the IR. The D5-brane embeddings exhibit nontrivial thermodynamics with multiple branches of solutions, whilst the thermal analogue of the interpolating D3-brane solution does not appear to exist.
 }
\begin{document}
\maketitle
\flushbottom
\section{Introduction}

The Polyakov/Wilson loop operators are  fundamental, gauge invariant, order parameters for confinement in Yang-Mills theories. They correspond to heavy quark probes of the gauge theory transforming in a given representation of the gauge group. Such probes can be viewed as impurities or point-like defects in the gauge theory. In the case of Yang-Mills theories with classical, holographic gravity/string theory duals \cite{maldacena, witten, gip}, a single heavy quark impurity in the fundamental representation maps to the end-point of a macroscopic string at the  conformal  boundary of the dual geometry \cite{malwil}. One may also consider  a collection of several such fundamental probes or, alternatively, probes transforming in various higher rank representations of the gauge group. At finite temperature, in theories with dual string/gravity descriptions, the Polyakov loops associated to such sources are given by string worldsheets and wrapped branes wound around the thermal cigar in a Euclidean black hole geometry \cite{Rey:1998bq, Witten:1998zw, fiol, us1, yamaguchi, Hartnoll:2006is}.

Our focus will be on specific D3- and D5-brane embeddings in AdS$_5 \times$ S$^5$ which {\it interpolate} between a heavy quark probe in a higher  rank symmetric or antisymmetric tensor representation  on the one hand, and multiple heavy quarks in the fundamental representation on the other. Such `flows' are controlled by the $0+1$ dimensional impurity theory on the heavy quark probe, interacting with the degrees of freedom of the $SU(N)$, ${\cal N}=4$ supersymmetric Yang-Mills (SYM) theory at large-$N$ and strong 't Hooft coupling.

The impurity theories for quark probes in generic tensor representations in ${\cal N}=4$ SYM were clarified  in \cite{passerini}, and the 0+1 dimensional theory for the antisymmetric representation at finite temperature was  solved in \cite{muck}\footnote{See closely related discussions in  \cite{kondo, Sachdev:2010uj}  within the context of holographic Kondo models.}.
The D3- and D5-brane embeddings we study in this paper, can be viewed as flows (in the renormalisation group sense) induced by an appropriate deformation of the impurity theories for symmetric and antisymmetric tensor representations.
We summarise our main findings below:
\begin{figure}
\begin{centering}
\includegraphics[width=4in]{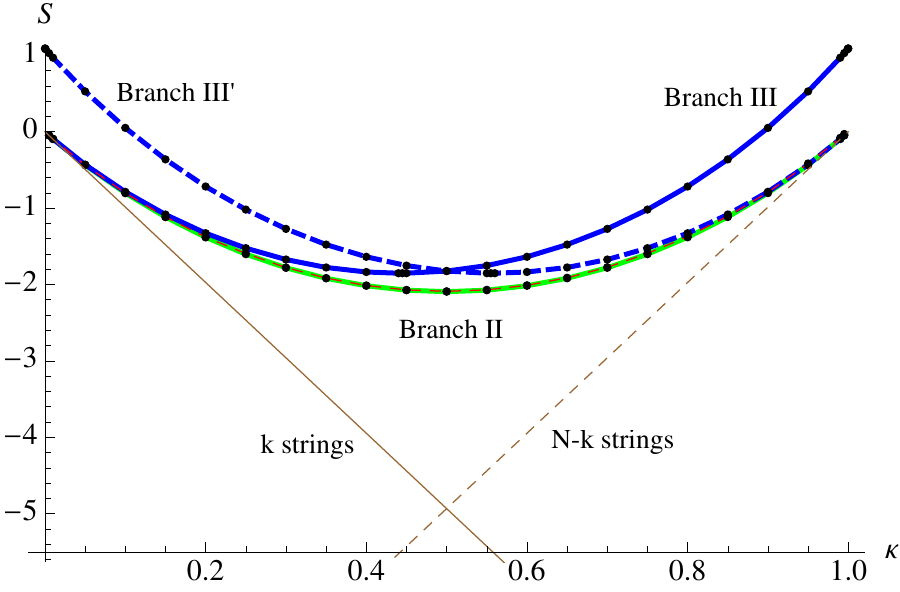}\\
\caption{\small The Euclidean actions for multiple branches of finite temperature D5-brane embeddings (at $T=\frac{3}{2\pi}$) corresponding in the infrared to a heavy quark impurity in the antisymmetric tensor representation, labelled by $\kappa\,\equiv\,\frac{k}{N}$. Branches II, III and III$^\prime$  represent  finite temperature flows from coincident quarks in the UV to the antisymmetric representation in the IR with $N$-ality $k$. The action for branch II is numerically indistinguishable from branch I -- the constant D5-brane embedding which computes the Polyakov loop in the antisymmetric representation.}
\label{D5}
\end{centering}
\end{figure}
\begin{itemize}
\item{
We obtain non-extremal  generalisations in global AdS$_5$ spacetime, of the zero temperature, supersymmetric D5-brane embeddings found originally in \cite{Callan:1998iq}. These interpolate between a D5-brane wrapped on a four-sphere inside the S$^5$ in AdS$_5\times$S$^5$ and a bundle or spike of fundamental strings approaching the boundary of AdS$_5$. We interpret the interpolating configuration as the  infrared screening of a collection of $k$ quarks in the fundamental representation, into the anti-symmetric tensor representation with the same $N$-ality. We confirm this interpretation by matching the UV and IR regimes of the flow with deformations of holographic  Wilson lines in fundamental and antisymmetric  representations, and by calculating the profile of the one-point function of the Lagrangian density sourced by the non-conformal impurity in the boundary CFT.}
\item{ Working in the global AdS$_5$-Schwarzschild black hole background, we identify {\em four} distinct branches (see figure \ref{D5}) of D5-brane embeddings. Three of these, labelled branches II, III and III$^\prime$ in figure \ref{D5}, represent flows from $k$ coincident heavy quarks to a source in the rank $k$ antisymmetric tensor representation. Branch I, which refers to the ``constant embedding'' yields the Polyakov loop in the antisymmetric representation and has been described in \cite{us1, yamaguchi, Camino:2001at}. Branches III and III$^\prime$ are exchanged under  a reversal of the string orientation and  $k\to N-k$ . On the other hand, branch II embeddings labelled by $k$ and $N-k$ have the same action and are thermodynamically degenerate with the constant embedding which has the lowest free energy (for non-collapsed solutions) for all temperatures.
 }
\item{The non-constant thermal D5-brane embeddings for a given $k$  exist below a critical temperature $T_c(\kappa)$ where $\kappa\equiv\frac{k}{N}$. At this temperature (for a given $\kappa$) branches II and III merge and disappear (figures \ref{d5merge} and \ref{Tcrit}). 
 Beyond a limiting high temperature, $T_c(\kappa=1)$ no non-constant finite temperature D5-brane embeddings appear to exist and the constant D5-brane embedding (branch I) remains the only nontrivial solution.
}
\item{ We find a new (BPS) D3-brane solution describing a flow from the Wilson line in the rank $k$ symmetric tensor representation in the UV to $k$ quarks in the fundamental representation in the IR.  From the AdS$_2$ asymptotics of the embedding we find that the flow is triggered by a VEV for a dimension one operator in the UV. Interestingly, the same interpretation appears for the D5-brane solution which interpolates between $k$ strings in the UV, and the antisymmetric representation in the IR. For both the D3- and D5-brane embeddings, the strength of static fields radiated by the corresponding impurity on the boundary {\em decreases} towards the IR, suggestive of a screening-like behaviour (figures \ref{interpolate} and \ref{d3vev}). 
The non-extremal or finite temperature generalisation of this embedding does not exist (for both planar and spherical horizon), which is consistent with the fact that the  expanded D3-brane solution of \cite{fiol} also appears not to exist in the AdS$_5$ Schwarzschild black hole background, as discussed in \cite{us1}.}
\end{itemize}

The paper is organised as follows: In section 2 we review the D5-brane action, its regularisation, the expanded D5-brane solutions related to holographic Wilson/Polyakov loops in the antisymmetric representation, and the interpretation of the interpolating D5-brane embedding as a flow.  In section 3,  the one-point function of the dimension four operator dual to the bulk dilaton sourced by the nonconformal D5-brane embedding. Section 4 discusses the results of the numerical analysis of non-extremal D5-brane flow embeddings, multiple branches and resulting thermodynamics. In Section 5, we review the D3-brane configurations that compute symmetric representation Wilson lines, and we present a different BPS solution and clarify its interpretation as an interpolating flow.


\section{Holographic Wilson/Polyakov loops}
Gauge theories with sufficient supersymmetries permit locally supersymmetric generalisations of Wilson loop observables. In the case of ${\cal N}=4$ SUSY Yang-Mills theory, the resulting Maldacena-Wilson loop \cite{malwil} along  some contour $C$, in a representation ${\cal R}$, involves a coupling to the six adjoint scalars of the ${\cal N}=4$ theory:
 \be
 W(C)\,=\,\frac{1}{\rm dim[{\cal R}]}\, {\rm Tr}_{\mathcal{R}} \mathcal{P} \exp \left[i \oint_C (A_{\a}\,\dot{x}^\a\,+\,i\,\P_I|\dot{x}|\,n^I(s))\,ds\right]\,,\quad I\,=\,1,\ldots 6\,.
 \ee
 Here $n^I$ is a unit vector in ${\mathbb R}^6$, picking out a direction locally in the internal space of the six adjoint scalars $\P_I$.  We will be interested in operators that preserve an $SO(5)$ subgroup of the full $SO(6)$  global symmetry acting on the six scalars, which is achieved by taking $n^I$ to be a constant along the contour $C$. Furthermore, we will take the contour $C$ to lie along the time direction, so that the Wilson line can be thought of as the world line of a heavy quark impurity. In the Euclidean formulation of the finite temperature theory, the contour winds around the thermal circle and yields the Polyakov loop in the representation ${\cal R}$.

The thermal state of the ${\cal N}=4$  theory on a spatial three-sphere, at large-$N$ and strong 't Hooft coupling, is given by string theory on the AdS$_5$-Schwarzschild black hole background. This assumes a temperature  $T$ above the Hawking-Page transition, when the finite volume theory is in the deconfined phase \cite{Witten:1998zw}. In the (Euclidean) AdS-Schwarzschild background\,,
\bea
&&ds^2\,=\,\left[\frac{dv^2}{f(v)}+ f(v)dt^2+v^2 d\Sigma_{\,3}^{\,2}\,+\,d\Omega_5^2
\right]\,,
\label{ads}\\\nonumber\\\nonumber
&&f(v)\,=\,1-\frac{v_+^2(1+v_+^2)}{v^2}+v^2\,,\qquad
v_+\,=\,\frac{1}{2}\left(\pi T\,+\sqrt{\pi^2 T^2\,-\,2}\right)\,.
\eea
the Polyakov loop is computed by the regularized action of string/brane embeddings which wrap the black hole cigar spanned by the $(t,v)$ subspace.

\subsection{Antisymmetric Wilson loops from D5-branes}
Wilson loops in the fundamental representation are computed by minimal area embeddings of open, fundamental string worldsheets in the dual geometry with the worldsheet boundary anchored to the contour $C$ on the conformal boundary of the dual gravity background.  Multiple coincident strings describe the insertion of a number $k$ of fundamental quarks. When the number of strings becomes large, the interactions between  them can cause the configuration to expand into suitably wrapped brane configurations with world-volume electric fields proportional to the number of quarks, or the $N$-ality of the representation in question \cite{fiol, us1, yamaguchi,  passerini}.

\subsubsection{Action and equations of motion}
\label{subsec:d5ac}
In order to compute the supersymmetric Wilson loop in the antisymmetric representation we consider a D5-brane wrapping an S$^4$ inside the S$^5$ in AdS$_5\times$S$^5$, thus preserving an $SO(5)$ subgroup of the full $SO(6)$ R-symmetry, whilst extending along the radial and temporal directions of the AdS$_5$ geometry. The symmetries of the configuration are those preserved by a point on the boundary three-sphere, simultaneously picking an orientation on the internal $S^5$.
The Dirac-Born-Infeld (DBI) action for the D5-brane is governed by the pullback of the background metric onto the worldvolume of the D5-brane embedding. As usual, we can use reparametrizations to pick the five world-volume coordinates to coincide with the spacetime coordinates $(t, v, \Omega_4)$, where $\Omega_4$ represents the coordinates of a four-sphere inside the $S^5$ in the geometry \eqref{ads}. Consistent with the $SO(5)$ symmetry of the configuration we can allow the polar angle $\theta$ specifying the location of the S$^4 \subset$ S$^5$, to depend on the radial coordinate $v$. Therefore, defining 
\be
d\Omega_5^2\,=\,d\theta^2\,+\,\sin^2\theta\,d\Omega_4^2\,,
\ee
the induced metric on the D5-brane worldvolume is
\be
{}^{\star}ds^2\,=\,\left[\frac{1}{f(v)}\,\left(\frac{\pa\,v}{\pa\,\sigma}\right)^2\,+\,\left(\frac{\pa\,\q}{\pa\,\sigma}\right)^2\right]\,d\sigma^2\,+\,f(v) dt^2\,+\,\sin^2\q \,d \Omega_4^2\,,
\ee
where $\sigma$ is a worldsheet coordinate which will eventually be set equal to $v$.
Formally, the DBI action along with the relevant  Wess-Zumino term is given by (in Euclidean signature)
\be
S_{\rm D5}\,=\,{\rm T}_{\rm D5}\int d\tau \,d^5\sigma\,e^{-\phi}\,\sqrt{{\rm det}\left({}^*g\,+\,2\pi\alpha'F\right)}\,-\,i g_s\,{\rm T}_{\rm D5}\int (2\pi\alpha'F)\wedge {}^*C_4\,+\,{S}_{\rm c.t.},\nonumber
\ee 
where $\phi$ is the background dilaton which can be set to zero in the AdS$_5\times$S$^5$ background dual to the conformal ${\cal N}=4$ theory.  We have also indicated the presence of counterterms $S_{\rm c.t.}$, required to regulate the formally divergent action. The D5-brane tension in terms of $N$ and the 't Hooft coupling $\lambda$ of ${\cal N}=4$ theory is\,:
\be
{\rm T}_{\rm D5}\,=\,\frac{N\sqrt{\lambda}}{8\pi^4}\,,\qquad
\lambda\,\equiv\,4\pi g_sN\,.
\ee
The embedding includes a non-vanishing worldvolume electric field along the radial direction which endows the configuration with $k$ units of string charge. In Euclidean signature this is a purely imaginary quantity
\be
iG\,\equiv\,2\pi\alpha'\,F_{tv}\,.
\ee
We will also need the pullback of the four-form potential, which is determined by the five form flux $F_5$, the latter being proportional to the volume-form on AdS$_5 \times$ S$^5$:
\be
C_4\,=\,\frac{1}{g_s}\left[\frac{3}{2}(\theta-\pi)\,-\,\sin^3\theta\,\cos\theta\,-\,\frac{3}{2}\sin\theta\cos\theta\right]\,{\rm Vol(S^4)}\,,\quad
F_5\,=\,\frac{4}{g_s}\sin^4\theta\,{\rm Vol(S^5)}\,.\nonumber
\ee
Here ${\rm Vol(S^4)}$ is the volume-form of the unit four-sphere.

The counterterms $S_{\rm c.t.}$ implement  boundary conditions on the D-brane action such that it is compatible with a Wilson loop in the boundary gauge theory. An open  string describing a holographic Wilson loop in 4D gauge theory must be subject to six Neumann transverse to the gauge theory directions and four Dirichlet boundary conditions along the gauge theory directions.  The basic DBI action is a functional of the embedding coordinates and conjugate momenta assuming we have Dirichlet boundary conditions for the variational problem. The fluctuations transverse to the four gauge theory directions must be exchanged for Neumann boundary conditions \cite{Drukker:1999zq, fiol}. This is implemented by performing a Legendre transform with respect to the boundary values of the coordinates and conjugate momenta excited in the brane embedding. In conjunction with this, we will also introduce a Lagrange multiplier constraint on the abelian field  which fixes the number of units of string charge carried by the configuration to $k$, so that
\be
S_{\rm c.t.}\,=\,S_{\rm UV}\,+\,S_{U(1)}\,,
\ee  
where,
\be
S_{\rm UV}\,=\, -\,\int_0^\beta dt\,\left[v\,\frac{\d S}{\d (\pa_\sigma v)}\,+\,(\q-\pi)\,\frac{\d S}{\d (\pa_\sigma \q )}\right]_{\rm AdS\, boundary }\label{legendre}
\ee
and
\be
S_{U(1)}\,=\, i\int dt\,d\sigma\,A_{\m}\,\frac{\d S}{\d A_{\m}}\,=\,- i\int  dt\,d\sigma\,F_{\m\n}\frac{\d S}{\d F_{\m\n}}\,=\,ik\int dt\,d\sigma\, F_{t\sigma}\,.
\ee
In the final step we have used Lagrange's equations for the gauge potential and  performed an integration by parts. The factor of $i$ is once again necessary to obtain the correct equations of motion in Euclidean spacetime.
 A byproduct of this treatment of boundary conditions is that it renders the on-shell action finite, providing a cut-off independent method of divergence regularisation. 

It is useful to define the quantity $D(\theta)$:
\be
D(\q)\,\equiv\,\sin^3\q\cos \q\,+\,\frac{3}{2}(\sin\q\cos \q-\q\,+\,\pi(1-\k))\,,\qquad \kappa\,\equiv\, \frac{k}{N}\,.\label{defd}
\ee
where, we are focussing on the limit $k,N\to\infty$ with $\kappa=k/N$ fixed. 
Finally, taking $\sigma=v$, we obtain an effective one-dimensional action for the D5-brane embedding in the (Euclidean) AdS-Schwarzschild background:
\be
S\,=\,{\rm T_{\rm D5}} \frac{8\pi^2}{3}\int_0^\beta dt\int_{v_+}^\Lambda dv\,\left[\sin^4\q\sqrt{1\,+\,f(v)(\pa_v \q)^2\,-\,G^2}\,-\,D(\q)\,G\right]\,+\,S_{\rm UV}\,.
\ee
The equation of motion for $G$ is algebraic whilst that for $\theta$ is a nonlinear second order system:
\bea
&&G\,=\,-\frac{D(\q)\,\sqrt{1\,+\,f(v)(\pa_v \q)^2}}
{\sqrt{D(\q)^2\,+\,\sin^8\q}}\label{defg}
\,,\\\nonumber\\\nonumber 
&&4\sin^4\q \left(\frac{\sin^3\q\cos\q}{D(\q)}-1\right)G\,=\,\frac{d}{dv}\left(\frac{D(\q)f(v)\pa_v \q}{G}\right)\,.
\eea
The equations are satisfied by two types of {\em constant} solutions, as explained in detail in \cite{us1}. The first of these is a collapsed solution,
where the angle $\q(v)\,=\,\pi$ (or $0$), and the D5-brane wraps a vanishing $S^4$. However, the presence of the electric flux, $G=+1$, renders this solution with a finite action which is naturally interpreted as the action for $k$ fundamental strings wound around the black hole cigar,
\be
S_{\rm collapsed}\,=\, -\frac{k\,\sqrt{\lambda}}{2\pi}\,v_+\,\beta\,.
\ee
In addition to these we also have non-trivial constant-$\theta$ solutions whose field theoretic interpretation is in terms of Polyakov loops in the antisymmetric tensor representation ${\cal A}_k$:
 \be
 S_{{\cal A}_k}\,=\,-\beta\,v_+\,\frac{N\sqrt{\lambda}}{3\pi^2}\,\sin^3\theta_k\,,\qquad
\pi(\k-1)\,=\,\sin\q_\k\cos\q_\k-\q_\k\,.
\label{constant}
\ee
The solution is invariant under the simultaneous transformation, $\k\to1-\k$ and $\q_k\to\pi-\q_k$, reflecting the charge conjugation property of the totally antisymmetric tensor representation. 

\subsection{BPS flow from $k$ quarks to the antisymmetric representation}
In \cite{Callan:1998iq}, a non-constant BPS solution for the DBI embedding action was found at zero temperature and in Poincar\'e patch of AdS$_5$ (planar conformal boundary) wherein $f(v)=v^2$. We will interpret that BPS solution as a flow between $k$ heavy quarks in the UV and an IR description given by the antisymmetric representation ${\cal A}_k$.

Let us first consider a small perturbation about the collapsed solution, $\theta=\pi$, discussed above and interpreted as a $k$-wound Polyakov loop:
\be
\theta(v)\,=\,\pi\,+\delta\theta(v)\,.
\ee
Linearising the  equation of motion for the perturbation, we obtain,
\be
v^2\,\delta\theta''(v)+2\,v\,\delta\theta'(v)\,=\,0\,.
\ee 
This can be  interpreted as  the equation for a massless scalar in AdS$_2$, and the general solution is a linear combination
\be
\delta\theta(v)\,=\,B\,-\,A\,v^{-1}\,,\label{uvfluct}
\ee
where the coefficients $B$ and $A$ have the usual interpretation as the source and a VEV for an operator with conformal dimension $\Delta\,=\,1$ in a dual conformal quantum mechanical system. 

For the constant solution which yields the Polyakov line in the antisymmetric representation ${\cal A}_k$, linearising the equation for a small fluctuation,
\be
\theta(v)\,=\,\theta_k\,+\,\delta\theta_k(v)\,,
\ee
the fluctuation satisfies the equation for a scalar with mass squared, $m^2=12$ in AdS$_2$:
\be
f(v)\,\delta\theta_k''(v)\,+\,f'(v)\,\delta\theta_k'(v)\,-\,12\,\delta\theta_k(v)\,=\,0\,,\qquad f(v)\,=\, v^2+\ldots
\ee
A scalar in AdS$_2$ is dual to an operator of conformal dimension $\Delta\,=\,\frac{1}{2}+\sqrt{\frac{1}{4}+m^2}$ which, in this case yields $\Delta=4$, an irrelevant deformation.
The asymptotic solutions to this equation of motion take the form,
\be
\q(v)\,=\,\tilde B\, v^3\,+\,\frac{\tilde A}{v^4}\,,
\ee
where, as usual, $\tilde B$ and $\tilde A$ represent the source and VEV respectively, for the dual operator. A deformation of the constant solution by this irrelevant mode will take the solution away from the antisymmetric representation in the UV i.e. as $v\to\infty$.
In particular, turning on a non-zero $\tilde B$  (but with $\tilde A=0$ so as to preserve the constant solution in the IR), and integrating outwards from the IR regime of small $v$, the equations of motion yield a power series
 \be
\q(v\to 0)\,=\,\q_\k\,+\,\tilde B\,v^3\,+\,2 \tilde B^2\,\cot\q_\k \,v^6\,+\ldots \label{IRexp}
\ee

A complete flow solution with this IR behaviour and the UV asymptotics of eq.\eqref{uvfluct} is captured by the zero temperature solution presented in \cite{Callan:1998iq}:
\be
v(\q)\,=\,\frac{A}{\sin\q}\left(\frac{\q-\sin\q\cos\q-\pi(1-\k)}{\pi\k}\right)^{1/3}\,,\label{nonconstant}
\ee
satisfying a first order BPS equation \cite{Callan:1998iq, Imamura:1998gk, Camino:2001at}.
In the limit $v(\theta)\to 0$, the numerator on the right hand side vanishes so that $\theta \to \theta_k$, whilst  the UV limit $v\to\infty$ corresponds to $\theta \to \pi$ or $0$ when the denominator of the right hand side vanishes.
The BPS solution describes a deformation of a collapsed UV solution with $\theta=\pi$, induced by a non-zero VEV, $A$, for an operator of scaling dimension $\Delta=1$, with no  non-normalizable mode or source. Its expansion in the IR matches \eqref{IRexp} with  
\be
\tilde B\,=\,\frac{\pi \kappa}{2 \sin\theta_k\,A^3}\,,
\ee
in line with the presence of an irrelevant deformation of the IR conformal impurity theory by a $\Delta=4$ operator.

One of our aims is to find numerical, finite temperature generalisations of the above flow and explore their thermodynamics. It will be useful to understand the computation of the regularised action for the flow solution.  The second order equation of motion at zero temperature, in the Poincar\'e  patch, is satisfied by solutions to the first order equation,
\be
\frac{\pa_\q v}{v}\,=\,\frac{\sin^5\q\,+\,D(\q)\cos\q}{\pa_\q(\sin^5\q\,+\,D(\q)\cos\q)}\,.\label{bps}
\ee
This BPS condition will allow us to analytically determine the action for the zero temperature flow configuration. We first rewrite the action as a functional of $v(\theta)$ and $v'(\theta)$:
\be
S\,=\,\frac{N\sqrt{\l}}{3\pi^2}\int dt\int_{\theta_k}^\pi d\q\,\left[\sin^4\q\,\sqrt{v^2\,+\,(\pa_{\q}v)^2(1-G^2)}\,-\,\partial_\theta v\,D(\q)G\right]\,+\,S_{\rm UV}\,.
\ee
Using the  expression for $G$ in \eqref{defg} and the first order condition
 \eqref{bps}, the first term above (the unregulated action) can be expressed as the integral over a total derivative,
\be
S\,=\,\frac{N\sqrt{\l}}{3\pi^2}\int dt\int_{\theta_k}^{\pi-A/\Lambda} d\q\,\pa_\q[v\,(\sin^5\q\,+\,D(\q)\cos\q)]\,+\,S_{\rm UV}\,.
\ee
The UV counterterm is determined by the Legendre transformed boundary 
condition \eqref{legendre} as\footnote{We drop the term $\sim (\pi-\theta)\frac{\delta S}{\delta(\partial_\sigma\theta)}$ which trivially vanishes at the boundary when $\theta$ approaches $\pi$.}
\be
S_{\rm UV}\,=\,-\frac{N\sqrt{\l}}{3\pi^2}\int dt\,v\,(\sin^5\q\,+\,D(\q)\cos\q)\left.\right|_{\theta\to\pi}\,.
\ee
This cancels off the divergent contribution from the UV, so the action of the flow solution is completely determined by the value of the boundary term in the IR, as  $v$ approaches zero, and $\theta \to \theta_k$. For the zero temperature embedding, this is also zero so the BPS solution has vanishing action as would be expected.

\section{Gauge theory VEV from interpolating D5-brane}
Static probes in the gauge theory  are sources for  Yang-Mills fields. A gauge-invariant description of these static fields is naturally provided by the IIB supergravity dual. A  heavy quark source or a Wilson line in the gauge theory induces a spatially varying expectation value for various Yang-Mills operators which can be inferred using bulk-to-boundary propagators in the AdS/CFT framework
\cite{Danielsson:1998wt, Callan:1999ki}. The simplest of these is the Lagrangian density $\sim \frac{1}{N}{\rm Tr}F_{\mu\nu}F^{\mu \nu}\,+\ldots$ which is dual to the dilaton field in the IIB supergravity dual \cite{Klebanov:1997kc}.
We will apply the results of
\cite{Callan:1999ki} to the non-constant D5-brane embedding which, as we have argued above, can be viewed as a flow within the $0+1$ dimensional impurity CFT. For a conformal probe the falloff of the expectation value of the static field sourced by it is a  power law dictated by the conformal dimension of the field. In the presence of the deformation discussed in the previous section however, the expectation value will acquire non-trivial  scale dependence which determines the length scale at which the $k$ heavy quarks get screened into the antisymmetric representation.

\subsection{The dilaton mode from the bulk}
In terms of gauge theory parameters, the relevant pieces in the 5D bulk supergravity action for the dilaton ($\phi$) in Einstein frame, are:
\be
S_\phi\,=\,- \frac{N^2}{8\pi^5}\,\int d^{10 }x\,\sqrt{-g}\,\,\tfrac{1}{2}g^{\mu\nu}\partial_\mu\phi\partial_\nu\phi \,+\,\int d^{10}x\, J_{\rm D5}(x)\,e^{\phi/2}\,.
\label{dilaton}
\ee
Here $J_{\rm D5}(x)$ is the effective Lagrangian density for the probe where the dependence on the dilaton has been factored out for later convenience. Explicitly,  in Einstein frame,
\bea
&&e^{\phi/2}\,J_{\rm D5}\,\equiv\,\\\nonumber
&&\qquad\frac{N\sqrt\l}{8\pi^4}\,\delta^{(3)}(\vec x)\,\delta(\q-\q(v))\,\left[\sin^4\q\sqrt{e^\phi\left(1\,+\,f(v)\,(\pa_v \q)^{2}\right)-G^2} \,-\,D G\right].
\eea
Notice that in order to identify the correct D5-brane action to which the dilaton couples, we have also included the term implementing the Legendre transform with respect to the auxiliary world-volume gauge field. The source is localized at the spatial coordinate
$\vec x=0$ in the gauge theory directions, and located at some internal polar angle $\theta(v)$ for any fixed $v$. Using the algebraic equation of motion for $G$ we rewrite the D5-brane action as,
\be
e^{\phi/2}J_{\rm D5}\,=\,
\frac{N\sqrt\l}{8\pi^4}\,\delta^{(3)}(\vec x)\,\delta(\q-\q(v))
\,e^{\phi/2}\,\sqrt{1\,+\,f(v)\,\q^{\prime 2}}\,\sqrt{D^2+ \sin^8\q}\,.
\ee
The form of the right hand side shows that it is natural to factor out the dependence on the dilaton as we have done in eq. \eqref{dilaton}. It also shows that in Einstein frame the dilaton couples to the wrapped D5-brane configuration in the same way as it does to the fundamental string.

In the classical supergravity limit $N\to \infty$, the self-interactions of the dilaton are suppressed by powers of $N^{-1}$, and it therefore suffices to focus on its linearized equation of motion. We further restrict attention to the zero mode of the dilaton on the S$^5$ to obtain the linearized 5D equation of motion, after integrating over the S$^5$ coordinates:
\be
-\frac{N^2}{8\pi^2}\,\partial_I\left(\sqrt{-g}g^{IJ}\partial_J\phi(\vec x, v)\right)\,=\,\frac{N\sqrt\l}{6\pi^2}\,\delta^{(3)}(\vec x)\,\sqrt{1\,+\,f(v)\,\q^{\prime 2}}\,\sqrt{D^2\,+\, \sin^8\q}\,.\label{linearized}
\ee
The indices $I,J$ represent the four {\em spatial} coordinates in AdS$_5$.  This equation  is readily solved via the 5D scalar retarded Green function $G_{\rm AdS}(x,x')$ in the AdS$_5$ geometry. For the situation without a black hole and in the Poincar\'e patch, the retarded propagator in empty AdS$_5$  was found in \cite{Danielsson:1998wt}. 
We will restrict attention to this situation, so that
\be
f(v)\,=\,v^2\,.
\ee
The scalar Green's function  is most compactly expressed in terms of the invariant timelike geodesic distance $s(x,x')$ in AdS$_5$ spacetime, defined as
\be
\cos s(x,x')\,=\,1+ \frac{(z-z')^2 + (\vec x\,-\, \vec x^{\,\prime})^2 -(t-t')^2}{2\,z\,z'}\,,\qquad z\,\equiv\,\frac{1}{v}\,.
\ee
The retarded propagator for a massless scalar in AdS$_5$ is then given by,
\be
G_{\rm AdS}(x,x')\,=\,-\frac{1}{4\pi^2\,\sin s}\,\frac{d}{ds}\left(\frac{\cos 2s}{\sin s}
\,\Theta\left(1-|\cos s|\right)\right)\,,
\ee
and the solution to the linearized equation eq.\eqref{linearized} is formally,
\be
-\frac{N^2}{8\pi^2}\,\phi(\vec x, v)\,=\,\frac{4\pi^2}{3}\int d^5x'\, G_{\rm AdS}(x,x')\,J_{\rm D5}(x')\,.
\ee
In order to extract the one-point function of the operator dual to the dilaton in the boundary gauge theory, we only need the leading term in the large-$v$ expansion of $\phi(\vec x, \,v)$. 
Plugging in the expression for $G_{\rm AdS}(x,x')$ the leading term in the large-$v$ expansion can be found following the steps outlined in
\cite{Danielsson:1998wt, Callan:1999ki}. We further rescale the dilaton so that its kinetic term is canonically normalized:
\be
\tilde\phi\,\equiv\,\frac{\sqrt{8\pi^2}}{N}\,\phi\,\simeq\,\frac{1}{v^4}\,\frac{5 \sqrt 2\,\sqrt\l}{16\pi^2}\,\int^\infty_{0} \frac{d\tilde v}{{\tilde v}^4}\,
\frac{\sqrt{1\,+\,v^2\,\q^\prime(\tilde v)^2}\,\sqrt{D^2\,+\,\sin^8\q}}{\left(\frac{1}{{\tilde v}^2}\,+\,|\vec x|^2\right)^{7/2}}\,.\label{canonical}
\ee
 Now we use the BPS D5-brane flow solution,
\be
v(\q)\,=\,\frac{A}{\sin\q}\left(\frac{\q-\sin\q\cos\q -\pi(1-\k)}{\pi \k}\right)^{1/3}\,,
\label{bpssol}
\ee
which interpolates between $\q=\pi$ as $v\to \infty$ and $\q\to \q_{\k}$ as $v\to 0$.  The dilaton is dual to the Lagrangian density of the boundary field theory. Correspondingly, we should see that the VEV of the dimension four operator 
\be
{\cal O}_{F^2}\,\equiv\,\tfrac{1}{N}\left({\rm Tr}\,F^2\,+ \ldots\right), 
\ee
sourced by the non-conformal impurity, exhibits a
non-trivial interpolation on the boundary between two qualitatively distinct behaviours.

\subsection{One-point function for ${\cal O}_{F^2}\sim\frac{1}{N}{\rm Tr}\,F^2\,+\,\ldots$}
The expectation value of the marginal operator ${\cal O}_{F^2}$ is given by the coefficient of the $v^{-4}$ term in the expansion of $\phi(v)$ above near the boundary. Accounting for the difference in normalization between $\tilde \phi$ and $\phi$, we obtain
\be
\left\langle {\cal O}_{F^2}\right\rangle\,=\, \frac{5\sqrt 2\sqrt\l}{16\pi^2}\int_{0}^{\infty}
\frac{d\tilde v}{{\tilde v}^4}\,
\frac{\sqrt{1\,+\,v^2\,\q^\prime(\tilde v)^2}\,\sqrt{D^2\,+\,\sin^8\q}}{\left({\tilde v}^{-2}\,+\,|\vec x|^2\right)^{7/2}}\,.
\ee
This expression encodes the complete position dependence of the VEV of the scalar glueball operator sourced by the impurity. It interpolates  between the short distance behaviour expected from $k$ heavy quarks in the fundamental representation, and for length scales  larger than a critical value determined by $A^{-1}$, it matches onto the antisymmetric tensor representation with $N$-ality $k$.

\begin{figure}
\begin{centering}
\includegraphics[width=4in]{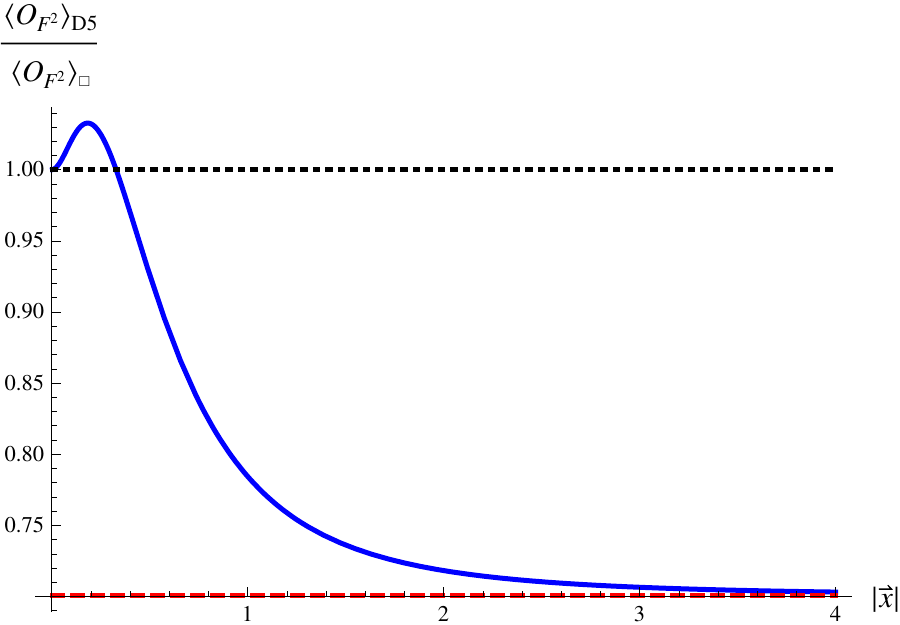}\\
\caption{\small The one-point function of ${\cal O}_{F^2}\sim \frac{1}{N}{\rm Tr}F^2$ for the nonconformal D5-brane embedding, plotted (solid blue) as a function of $|\vec x|$, for $\kappa\,=\,\frac{k}{N}\,=\,0.2$ and a VEV in the UV impurity theory, $A=2.0$. Dividing by the one-point function for $k$ fundamental quarks, the resulting curve interpolates between unity (dotted black) for $k$ fundamental quarks and $\frac{2}{3\pi\kappa}\sin^3\theta_k$ (dashed red) for the antisymmetric representation of rank $k$.}
\label{interpolate}
\end{centering}
\end{figure}

The integral above is not analytically tractable, and can be evaluated numerically. However we can analytically obtain its asymptotics, both for small and large $|\vec x|$ in the zero temperature BPS situation.  We first define a dimensionless integration variable
$\alpha = v |\vec x|$, in terms of which the one-point function is, 
\be
\left\langle{\cal O}_{F^2}(\vec x)\right\rangle\,=\, \frac{5\sqrt 2}{16\pi^2}\,\frac{\sqrt\l}{|\vec x|^4}
\,\int_0^\infty d\a\,\frac{\a^3}{(1+\a^2)^{7/2}}
\sqrt{1+\a^2\,\left(\tfrac{d\q}{d\a}\right)^2}
\sqrt{D^2\,+\,\sin^8\q}
\ee
where $\q$ is a function of $v = \a/|\vec x|$, defined implicitly
in eq.\eqref{bpssol}. For any fixed $\a$, we therefore have:
\bea
&&\q(\a/|\vec x|)\,\to\,\pi\,\qquad {\rm for}\quad |\vec x| \to 0\,,\\\nonumber\\\nonumber
&&\q(\a/|\vec x|)\,\to\,\q_\k\,\qquad {\rm for}\quad |\vec x| \to \infty\,.
\eea
This means that for large and small $|\vec x|$ the internal angle $\theta$ approaches constant values. Importantly, in each of these limits, the scalar glueball VEV is proportional to $|\vec x|^{-4}$, as expected from dimensional analysis for a conformal probe, but with a different normalisation constant.
In both the asymptotic regimes of small and large $|\vec x|$ when $\theta$ approaches a constant, the integral simplifies considerably. The integration over $\alpha$ yields,
\be
\int_0^\infty d\a\,\frac{\a^3}{(1+\a^2)^{7/2}}\,=\,\frac{2}{15}\,.
\ee
This is multiplied by  $\sqrt{D^2 + \sin^8\theta}$  evaluated at $\theta=\pi$ or $\theta=\theta_k$ and we obtain,
\bea
\left\langle{\cal O}_{F^2}\right\rangle\,&=&\,\frac{\sqrt 2}{24\pi^2} \left(\frac{3\pi\k}{2}\right)\,\frac{\sqrt\l}{|\vec x|^4}\,,\qquad|\vec x|\quad {\rm small}\,,\label{f^2D5}\\\nonumber\\\nonumber
&=&\,\frac{\sqrt 2}{24\pi^2} \left(\sin^3\q_k\right)\,\frac{\sqrt\l}{|\vec x|^4}\,,\qquad|\vec x|\quad {\rm large}\,.
\eea
Using $\kappa=k/N$, we conclude that the one-point function of the glueball operator interpolates between that corresponding to a bundle of $k$ coincident quarks each in the fundamental representation, and that for a collection of $k$ quarks transforming in the antisymmetric tensor representation. The full interpolating function is plotted numerically in figure \ref{interpolate}.
Since $\sin^3\theta_\kappa < \frac{3\pi}{2}\kappa$ for all $\kappa$, the figure displays the ``screening'' of  $k$ coincident quarks into the rank-$k$ antisymmetric tensor representation. The screening is induced by a VEV, or normalizable mode in the effective AdS$_2$ geometry induced on the D5-brane embedding in the UV regime.

\section{Finite temperature D5-brane embeddings}
It is remarkable that at zero temperature there exist non-constant D5-brane embeddings (preserving $SO(5)$ global symmetry) which are BPS and therefore degenerate with the constant embeddings with $\theta=\theta_k$ or $\theta=\pi$.
We would like to know what happens to such embeddings away from extremality, i.e. when the dual gauge theory is in a thermal state. In this section we perform a numerical investigation of the thermodynamics of non-constant solutions.  In order to explore both high and low temperatures we will work in the global AdS-Schwarzschild background which corresponds to the thermal ${\cal N}=4$ theory on a spatial three-sphere.

We will first search for non-extremal D5-brane flow embeddings, and the 
corresponding value of the deformation in the UV regime, given by the integration constant $A$ (see eq.\eqref{uvfluct}), for which the embeddings exist at a given temperature. Once such embeddings are identified we will investigate their thermodynamics and how they compete in the thermal ensemble with the collapsed $k$-quark solution and the constant antisymmetric tensor embedding.

Our main result is that we find three classes of non-constant embeddings interpolating between  $k$-quarks and the rank $k$ antisymmetric representation. The actions of two of these categories are exchanged under the action $k\to N-k$, whilst the third is symmetric under this operation which acts as charge conjugation. The free energy of this latter class of solutions is numerically indistinguishable from the constant antisymmetric tensor embedding, for any temperature. It is curious that expectation value of the short screening length flow coincides with that of the antisymmetric Polyakov loop, suggesting some underlying analytic method for its evaluation, despite the fact that conformal invariance is broken.
The different classes of solutions also merge with each other at some critical temperature for a given value of the parameter $\kappa=\frac{k}{N}$.

\subsection{Numerical solutions}
At finite temperature conformal invariance and supersymmetry are broken, and therefore no BPS condition exists. The full second order equation of motion for the polar angle $\theta(v)$, characterising our D5-brane embedding in the AdS-Schwarzschild background is,
\be\
4\sin^4\q \left(\frac{\sin^3\q\cos\q}{D(\q)}-1\right)G\,=\,\frac{d}{dv}\left(\frac{D(\q)\pa_v \q}{G}\left(1-\frac{v_+^2(1+v_+^2)}{v^2}+v^2\right)\right)\,.
\ee
The auxiliary field $G$ and the function $D(\theta)$ are defined as before in eqs.\eqref{defd} and  \eqref{defg}. The resulting nonlinear differential equation is solved with the UV boundary conditions appropriate for describing the collapsed solution \eqref{uvfluct} with $k$ coincident fundamental strings (quarks). Specifically, we build the large $v$ expansion for some deformation $A$, with fixed $\kappa$ and temperature $T$ (or $v_+$) :
\bea
\q(v)&&\left.\right|_{v\to\infty} \,=\\\nonumber
&&\,\pi-\,\frac{A}{v}\,+\,\frac{A(2-A^2)}{6\,v^3}\,+\,\frac{2A^4}{9\pi\k v^4}\,-\,\frac{A(3A^4-8A^2+8(1\,+\,v_+^2\,+\,v_+^4))}{40 v^5}+\ldots
\eea
Using this expansion, we solve the equation of motion numerically, integrating 
in towards the horizon. Acceptable solutions are those which extend all the way to the horizon, remaining smooth and finite for $v_+ \leq v < \infty$. Once a solution is found by integrating in from the UV, we check the same by integrating outwards from the horizon towards the UV. For the latter procedure, we begin with a general near horizon expansion of the form,
\be
\q(v)\left.\right|_{v\to v_+}\,=\,\q_+\,+\,g_1(\q_+,\k ,v_+)\,(v-v_+)\,+\,g_2(\q_+,\k ,v_+)(v-v_+)^2\,+\,\ldots
\ee
and the full solution is determined numerically. The numerical solutions and their actions presented here were evaluated using a (large) radial UV cutoff $\Lambda\sim 10^9$, and the results  were found to be stable against changes in $\Lambda$ (for large enough $\Lambda$).

At generic temperatures below a critical value (for fixed $\kappa$), which we refer to as $T_c(\kappa)$, we find three non-constant embeddings. Together with the two constant embeddings -- the collapsed $k$-quark solution and the expanded antisymmetric tensor embedding, we therefore have five branches of solutions. The branches are each distinguished by the values of the UV deformation parameter $A$ associated to each of them, and by their (regulated) Euclidean action/free energy. We label the constant solution for the antisymmetric representation as ``branch I" and classify the non-constant embeddings as follows (see figure \ref{branches}):
\begin{figure}
\begin{center}
\includegraphics[width=2.5in]{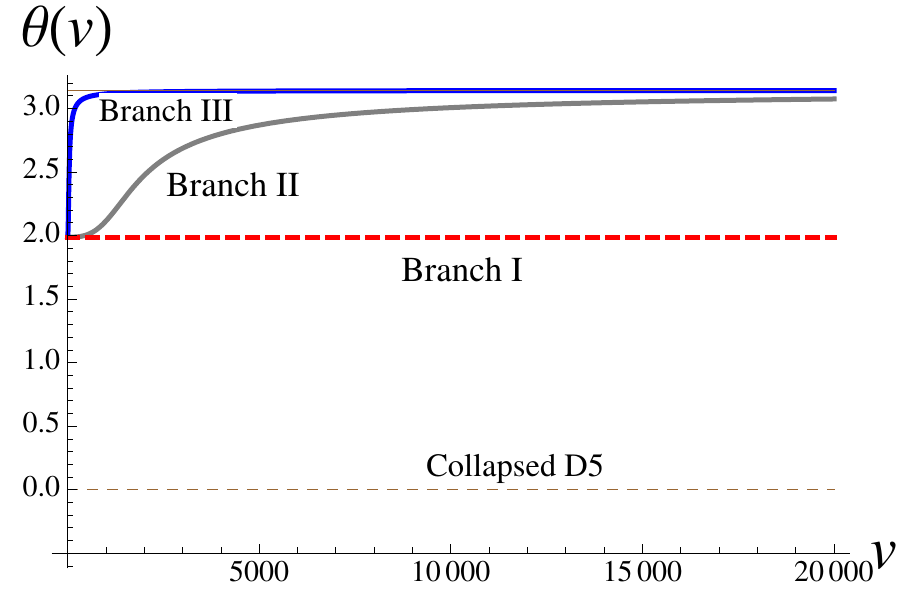}\hspace{0.8in}\includegraphics[width=2.5in]{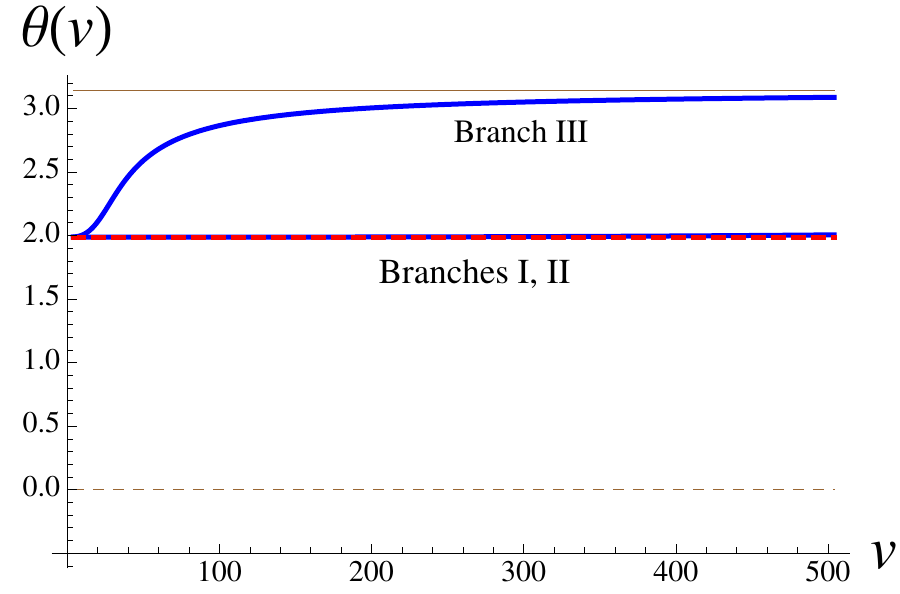}
\caption{\small The polar angle $\theta(v)$  of the ${\rm S}^4 \subset {\rm S}^5$ wrapped by the D5-brane, as a function of the radial coordinate, for  
different branches of D5-brane embeddings, with $\kappa=0.25$ and  $T\approx 1.313$ (or $v_+\,=\,4$). Upon zooming in near the horizon (right) we see that Branch III approaches the black hole horizon smoothly.}
\label{branches}
\end{center}
\end{figure}
\begin{enumerate}
\item{The solutions labelled as branch II in figure \ref{branches} start off near $\theta=\pi$ at the conformal boundary $(v\to \infty)$ and approach the value $\theta_k$ at the horizon $v=v_+$. The value of the angle $\theta_k$ (eq.\eqref{constant}) at the horizon characterises the constant antisymmetric tensor embedding ${\cal A}_k$. We find numerically that the action for this family of solutions is symmetric under the exchange $\kappa\to 1-\kappa$, as displayed in figure \ref{D5} and in addition, is numerically indistinguishable from the action \eqref{constant} for the constant embeddings \cite{us1}.
}
\item{We find another qualitatively distinct family of non-constant solutions which we label branch III. These solutions approach $\theta=\theta_+\approx \theta_k$ at the horizon, with the transition scale (screening length) to the antisymmetric representation deeper in the infrared relative to a  branch II solution for the same value of $\kappa$, as  evident in figure \ref{branches}. Furthermore, the action for this family of embeddings is not symmetric under $\kappa \to 1-\kappa$. }
\item{For a given value of $\kappa$, we define a third family branch II\'I, of non-constant embeddings obtained by taking a solution from branch III with $\kappa^\prime\,=\, 1-\kappa$ and $G\to -G$ (or $\theta\to \pi-\theta$). This operation reverses the string orientation and describes $N-k$ anti-quarks being screened to a rank $N-k$ antisymmetric tensor  representation $\bar{\cal A}_{N-k}$ of anti-quarks. The latter has the same $N$-ality as $k$-quarks in the antisymmetric tensor representation ${\cal A}_k$. }
\end{enumerate}
\begin{figure}
\begin{center}
\includegraphics[width=2.5in]{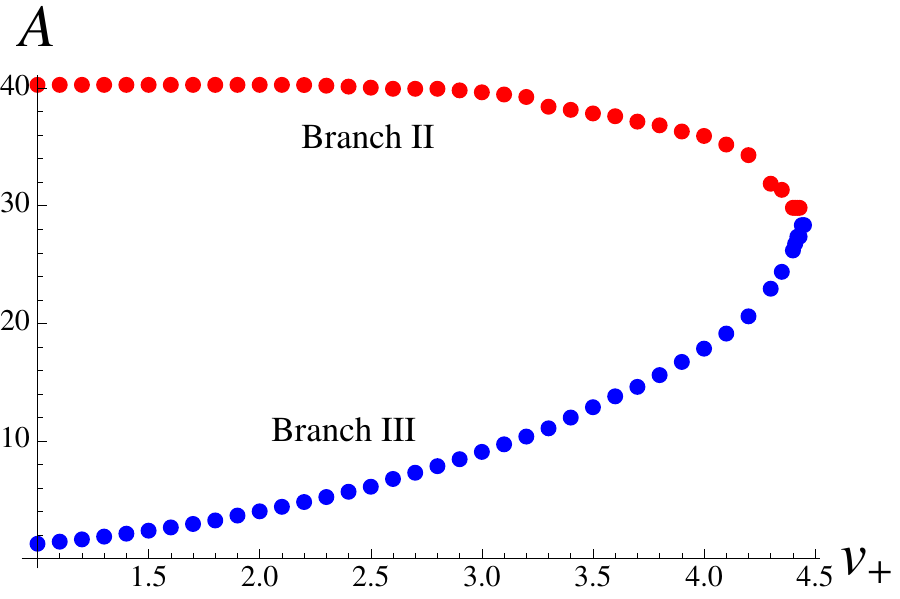}\hspace{0.5in}\includegraphics[width=2.5in]{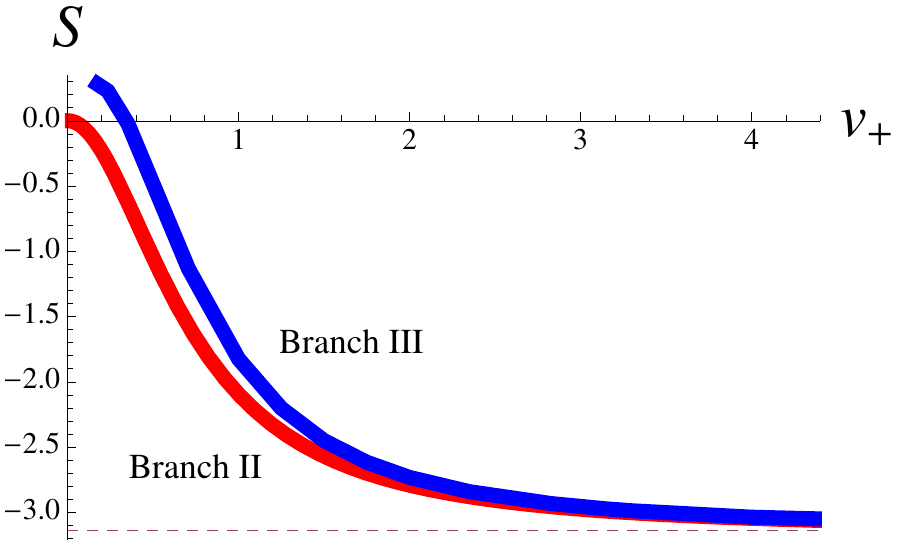}
\caption{\small Branches II and III merge at a critical temperature $T_c(\kappa)$ (for a given $\kappa$), beyond which interpolating embeddings do not appear to exist. {\bf Left:} The value of the UV deformation (VEV) as a function horizon size (temperature) for the two branches for $\kappa=0.25$. {\bf Right:} Actions for the two interpolating solutions merging as a function of temperature (horizon size).}
\label{d5merge} 
\end{center}
\end{figure}
\begin{figure}
\begin{center}
\includegraphics[width= 2.5in]{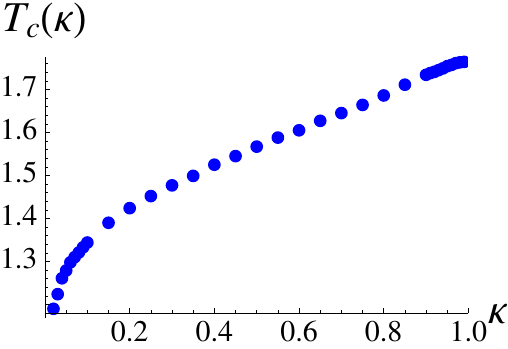}
\caption{\small Critical temperature $T_c(\kappa)$ at which branches II and III of interpolating D5-brane solutions merge.}\label{Tcrit}
\end{center}
\end{figure}
The actions for all non-constant embeddings are evaluated using the UV counterterms \eqref{legendre}. We find that the non-constant flow solutions display nontrivial thermodynamics as a function of temperature. As the temperature of the AdS-Schwarzschild background is increased, we find that branches III and II\'I merge with branch II as displayed in figure \ref{d5merge}. The critical temperature at which this merger occurs depends on $\kappa$, and  appears to increase monotonically with $\kappa$ (figure \ref{Tcrit}).
Beyond the critical temperature $T_c(\kappa)$ for any given $\kappa$, integrating in from the UV with $\theta(v\to \infty)\,=\,\pi$ we have been unable to unambiguously identify non-constant solutions that smoothly approach the horizon. We conclude that D5-brane embeddings interpolating between $k$ coincident strings in the UV and the antisymmetric representation in the IR, cease to exist for temperatures above $T_c(\kappa)$.

\section{Symmetric Wilson loops and D3 branes}
Wilson loops in the  totally symmetric representation of rank $k$ are computed by D3-branes with worldvolume 
AdS$_2\times {\rm S}^2 \subset {\rm AdS}_5$ 
\cite{fiol, passerini, Hartnoll:2006is}. The induced metric on the D3-brane worldvolume yields a constant size $S^2$ with radius given by $\tilde \kappa\,=\,\frac{k\sqrt\lambda}{4N}$. 
In this section we will describe a (BPS) solution which generalises the D3-brane solution of \cite{fiol} so that the embedding represents a flow from a source in the symmetric representation  in the UV to a collection  of $k$ quarks (coincident strings) in the IR. We will confirm this interpretation by computing the VEV of the glueball operator in the gauge theory, and we will again find an interpretation indicative of a screening effect.

\subsection{D3-brane flow solution}
A point in ${\mathbb R}_{3}$ preserves an $SO(3)$ rotational symmetry, while a choice of internal orientation of the BPS Wilson loop breaks the global $SO(6)$ R-symmetry to $SO(5)$. The D3-brane embedding preserves the same symmetries, and whilst sitting at a point on the $S^5$ of the dual background, wraps an $S^2$  centred at the origin along the spatial ${\mathbb R}^3$ slices of AdS$_5$. Writing the AdS$_5$ metric in  Poincar\'e patch  as:
\be
ds^2\,=\,\frac{dz^2}{z^2}\,+\,\frac{1}{z^2}\left(-dt^2\,+\,d\rho^2\,+\,\rho^2\,d\Omega^2_2\right)\,,
\ee
the pullback metric on the D3-brane worldvolume is,
\be
{}^*ds^2\,=\,\frac{d\sigma^2}{z^2}\,\left[\left(\frac{\partial \rho}{\partial\sigma}\right)^2\,+\left(\frac{\partial z}{\partial \sigma}\right)^2\right]\,+\,\frac{1}{z^2}\left(-dt^2\,+\,\rho^2d\Omega_2^2\right)\,.
\ee
As in the case of the D5-brane solution, a worldvolume electric field $F_{t\sigma}$ is also switched on to generate $k$ units of fundamental string charge. The Wess-Zumino  term on the D3-brane worldvolume is determined by the pullback of the RR four-form potential which follows directly from the five-form flux proportional to the volume form of AdS$_5$:
\be
{}^*C_4\,=\,-\frac{i}{g_s}\,\frac{\rho^2}{z^4}\,\partial_\sigma\rho\,dt\wedge d\sigma\wedge {\rm Vol}\left({\rm S}^2\right)\,.
\ee
The DBI and WZ terms of the D3-brane action together yield:
\be
S_{\rm D3}\,=\, 4\pi\,{\rm T}_{\rm D3}\int dt \,d\sigma\, \frac{\rho^2}{z^4}\left(\sqrt{(\pa_\sigma \rho)^2\,+\,(\pa_\sigma z)^2\,-\,G^{\,2}\,z^{\,4}}\,-\, \pa_\sigma \rho\right)\,+\,S_{\rm c.t.}\,.\label{d3dbi}
\ee
The D3-brane tension ${\rm T}_{\rm D3}\,=\,\frac{N}{2\pi^2}$ and $G\,\equiv\,2\pi\alpha'\,F_{t\sigma}$, while the counterterms are determined by the Legendre transform procedure described in \cite{fiol, us1} and in section \ref{subsec:d5ac}. In particular, the counterterms enforce the condition that the D3-brane is endowed with $k$ units of string charge:
\be
S_{\rm c.t.}\,=\,S_{U(1)}\,+\,S_{\rm UV}\,,\qquad\qquad S_{U(1)}\,=\,-
k\,\int dt \,d\sigma\, F_{t\sigma}\,,
\ee
and exchanging Dirichlet for Neumann boundary conditions generates boundary  counterterms,
\be
S_{\rm UV}\,=\,-\int dt\,\left[\rho\,\frac{\d\,S}{\d\,(\pa_\sigma \rho)}\,+\,z\,\frac{\d\,S}{\d\,(\pa_\sigma z)}\right]_{\rm UV}\,.
\ee
Analogous to the D5-brane case the equations of motion yield a nonlinear system, with the electric field being determined algebraically,
\be
G\,=\,\tilde\k\,\frac{\sqrt{(\pa_\sigma \rho)^2\,+\,(\pa_\sigma z)^2}}{\sqrt{\rho^4\,+\,\tilde\k^2\,z^4}}\,,\qquad\qquad\tilde\kappa\,\equiv\,\frac{k\sqrt\lambda}{4N}\,.
\ee
Picking the gauge $\sigma\,=\,z$, the equation of motion for $\rho(z)$ becomes:
\be
\frac{2\,\rho}{z^4}\left(\frac{\rho^2\,G}{\tilde \k}\,-\,\pa_z\rho\right)\,=\,\frac{d}{dz}\left(\frac{\tilde \k\,\pa_z\rho\,-\,\rho^2\,G}{z^4\,G}\right)\,.
\ee
The equations of motion are satisfied by configurations that solve the first order equation
\be
\frac{\pa\, \rho}{\pa\, z}\,=\,G\,\frac{\rho^2}{\tilde\k}\,,
\ee
which we can view as a  BPS condition. The most general solution to the first order equation of motion is,
\be
G\,=\,\frac{1}{z^2}\,,\qquad\qquad \rho\,=\,\frac{z\,\tilde \k}{1+a\,z\,\tilde\k}\,,\label{d3flow}
\ee  
where $a$ is the constant of integration. We will only consider the case $a>0$ in this paper. Interestingly, the same solution with $a<0$ has appeared in \cite{schwarz} in a different context relevant for describing solitonic lumps on probe D3-branes\footnote{When $a<0$, $\rho$ diverges at some fixed $z=1/|\tilde\kappa a|$, and therefore the configuration describes the Coulomb phase of ${\cal N}=4$ theory with $U(N+1)\to U(1)\times U(N)$, with a solitonic lump dual to the  funnel-shaped spike in the D3-brane probe representing the $U(1)$ factor.}.
The AdS$_2 \times {\rm S}^2$ embedding of \cite{fiol} describing the straight BPS Wilson line in the symmetric representation is obtained by setting $a=0$. On the other hand, taking the  limit of large $a$ we obtain the trivial or collapsed embedding:
\be
\lim_{a\to\infty}\rho(z)\,=\,0\,.
\ee
The collapsed D3-brane solution represents a bundle of $k$ coincident strings, since the action (omitting the boundary counterterms) is then simply
\be
S_{\rm D3}\left.\right|_{\rm collapsed}\,=\,\int dt\int d(z^{-1})\,\frac{k}{2\pi\alpha'}\,,
\ee
representing the tension of $k$ strings oriented along the radial direction of the AdS geometry. As we will explain in more detail below, the solution \eqref{d3flow} interpolates between the symmetric representation D3-brane and $k$ strings.
\subsection{Regularized action}
The general solution \eqref{d3flow} satisfies a first order equation and therefore, as is generally the case for (straight) BPS Wilson lines, we expect that it has vanishing action. For the BPS solution with $G\,=\,z^{-2}$ (in the gauge $\sigma=z$), the DBI and Wess-Zumino terms cancel each other off, leaving only the counterterms. Evaluating these separately with a boundary cutoff
$z\,=\,\varepsilon\,\ll 1$, we find,
\bea
&& S_{\rm U(1)}\,=\,-\frac{2N}{\pi}\int dt\int_\infty^\varepsilon dz\,\tilde \kappa \,G\,=\,\frac{2N}{\pi}\int dt\,\frac{\tilde\k}{\varepsilon}\,,\\\nonumber\\\nonumber
&& S_{\rm UV}\,=\,-\frac{2N}{\pi}\int dt\left[\rho\left(\frac{\pa_z\rho\,\tilde\k\,-\,\rho^2\,G}{z^4\,G}\right)\,+\,\frac{\tilde \k}{z^3\,G}\right]_{z=\varepsilon}\,=\,-\frac{2N}{\pi}\int dt\,\frac{\tilde\k}{\varepsilon}\,.
\eea
Therefore, the counterterms also sum to zero and the worldvolume action vanishes.
It is curious to note that despite the scale dependence of the flow solution, neither the world sheet $U(1)$ field nor the action depends upon the scale $a$. 

\subsection{Interpretation as a flow}
For positive values $a>0$, we may interpret the solution \eqref{d3flow} as a flow, interpolating between the symmetric representation Wilson loop in the UV, 
\be
\rho(z)\left.\right|_{z\to 0}\,=\,\tilde\kappa\, z\,-\,a\,\tilde\kappa^2\,z^2\,+\ldots\label{uvd3}
\ee
and a ``spike'' or bundle of $k$ strings in the IR regime (large $z$):
\be
\rho(z)\left.\right|_{z\gg 1}\,=\,\frac{1}{a}\,-\,\frac{1}{a^2 \,z \,\tilde\kappa}\,+\,\ldots\,.\label{ird3}
\ee
The interpretation of the IR spike as a bundle of $k$ fundamental strings is not automatically clear from the solution itself or from the Lagrangian density
\eqref{d3dbi} since the DBI and Wess-Zumino terms cancel each other out for any value of $z$. In order to arrive at the correct physical picture we first examine the induced metric on the D3-brane embedding, the behaviour of fluctuations about the symmetric and collapsed embeddings, and finally the behaviour of the scalar glueball operator sourced by the impurity in the boundary ${\cal N}=4$ theory.
The induced metric on the D3-brane worldvolume for any non-vanishing deformation $a$ is,
\be
{}^{\star}ds^2\,=\,\frac{1}{z^2}\left[\left(1+\frac{\k^2}{(1\,+\,a\,z\,\k)^{\,4}}\right)dz^2\,+\,dt^2\right]\,+\,\frac{\k^2}{(1\,+\,a\,z\,\k)^{\,2}}d\Omega_{\,2}^{\,2}\,.
\ee
When $a=0$ this reduces to AdS$_2\times{\rm S}^2$ where the AdS$_2$ has radius $\sqrt{1+\kappa^2}$ and the S$^2$ radius is $\kappa$. In the limit $a\to\infty$ (which can also be viewed as a large-$z$ or IR limit) we have,
\be
{}^*ds^2\left.\right|_{a\gg 1}\,=\,\frac{1}{z^2}\,\left[\left\{1+ {\cal O}\left((a z)^{-4}\right)\right\}\,dz^2\,+\,dt^2\right]\,+\,\frac{1}{a^2 z^2}\,d\Omega_2^2\,.
\ee
In this limit, the  $S^2$ shrinks and we may view this as the approach towards the collapsed D3-brane solution which describes a bundle or spike  of $k$ fundamental strings.

\subsection{Fluctuations about UV and IR regimes}
The only mode excited in our D3-brane flow solution is the radius of the $S^2$ wrapped by the brane, which is a scalar from the point of view of the induced AdS$_2$ on the worldvolume of the embedding \footnote{For a general analysis of the spectrum of fluctuations about D3- and D5-brane embeddings computing higher rank Wilson loops, we refer the reader to the works \cite{Faraggi:2011bb, Faraggi:2011ge, Faraggi:2014tna}.}. Let us separately consider the linearised fluctuation of this radial mode about the expanded D3-brane solution with $a=0$ and the collapsed embedding. Therefore, we write 
\be
\frac{\rho}{z}\,=\,\tilde\kappa\,+\,\delta(z)\,,
\ee
for the solution corresponding to the symmetric representation Wilson line, and  
\be
\frac{\rho}{z}\,=\,\delta(z)\,,
\ee
for the collapsed D3-brane embedding. The linearised equation of motion for $\delta(z)$ in each case is then:
\bea
&&{\rm  Symmetric:}\quad \delta''(z)\,=\,0\,, \qquad\qquad\,\delta(z)\,=\,B\,-\, A\, z\,.\label{uvirfluct}
\\\nonumber\\\nonumber
&&{\rm   Collapsed:}\quad z^2\delta''(z)\,-\,2\delta(z)\,=\,0\,,\qquad\qquad
\delta(z)\,=\,\tilde A\, z^2\,+\,\frac{\tilde B}{z}\,.
\eea
The first is the equation of motion for a massless scalar in AdS$_2$, dual to a 
$\Delta=1$ operator in the dual conformal quantum mechanics describing the impurity. The second equation corresponds to an AdS$_2$ scalar of mass $m^2=2$ which in turn is dual to an (irrelevant) operator of dimension $\Delta\,=\, 2$ using the standard relation $\Delta\,=\,\frac{1}{2}+\sqrt{m^2+\frac{1}{4}}$.

Let us now compare the expansions \eqref{uvd3} and \eqref{ird3} for $\rho(z)$ in the UV and IR with the linearized fluctuations in \eqref{uvirfluct}. Noting that fluctuation in $\rho(z)$ is actually given by $z\,\delta(z)$, we immediately infer that,
\be
\tilde A\,=\,B=\,0\,,\qquad A\,=\,a\,\tilde\kappa^2\,,\qquad \tilde B\,=\,
\frac{1}{a}\,.
\ee
Therefore the UV regime corresponds to the expanded D3-brane deformed by a VEV, $A\neq 0$ for a $\Delta=1$ operator, whilst the IR regime can be viewed as a deformation of the collapsed configuration by a source $\tilde B\,=\,\tilde\kappa^2/A$ for an  irrelevant operator. The resulting physical picture is therefore remarkably  similar to what we found for the D5-brane flow between $k$ strings in the UV and the expanded D5-brane in the IR and which could be interpreted as a screening effect in the dual gauge theory. In order to understand whether the symmetric representation source is really ``screened'' to yield $k$ coincident quarks, we now turn to the computation of the scalar glueball VEV in the ${\cal N}=4$ theory.

\subsection{One-point function for ${\cal O}_{F^2}\sim\frac{1}{N}{\rm Tr}F^2\,+\,\ldots$}
\begin{figure}
\begin{centering}
\includegraphics[width=4in]{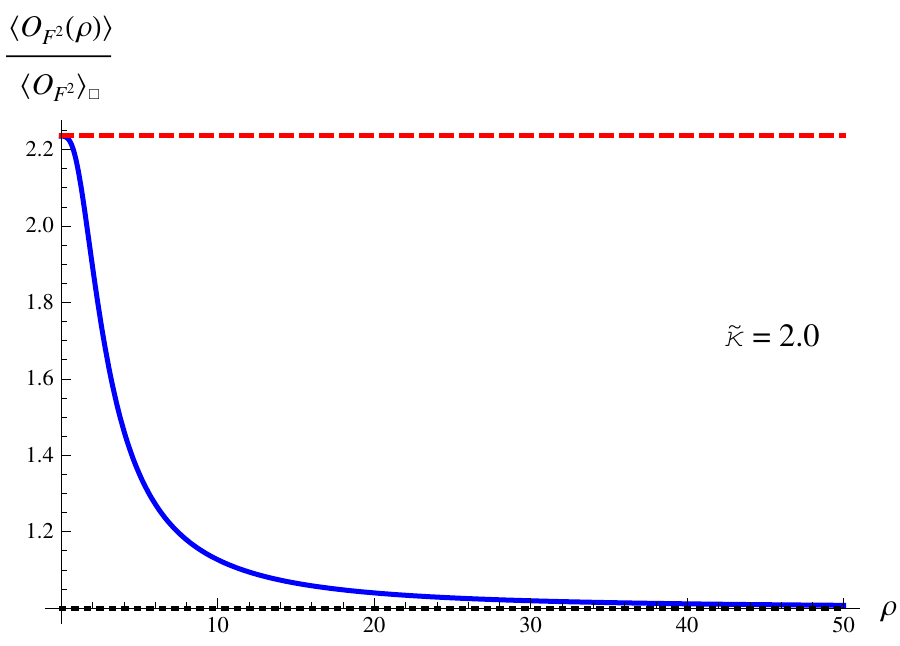}\\
\caption{\small One-point function of ${\cal O}_{F^2}\sim \frac{1}{N}{\rm Tr}F^2$ for the nonconformal D3-brane embedding, plotted (solid blue) as a function of $\rho=|\vec x|$, for $\tilde \kappa\,=\,\sqrt\lambda\frac{k}{4N}\,=\,2.0$ and deformation parameter/VEV in the UV impurity theory, $a=0.3$. Normalizing with respect to the one-point function for $k$ fundamental quarks, the curve interpolates between $\sqrt{1+\kappa^2}=2.236$ in the UV corresponding to the symmetric representation, and the fundamental representation in the IR.}
\label{d3vev}
\end{centering}
\end{figure}
As in the D5-brane case, we will now deduce  the one-point function for the marginal operator ${\cal O}_{F^2}$ dual to the dilaton in the bulk AdS$_5$ geometry sourced by our D3-brane flow solution. The details of the analysis proceed similarly to the D5-brane example, with the one key difference that the D3-brane embedding resides entirely in the non-compact AdS$_5$ directions and is point-like on the internal S$^5$. The corresponding calculation for the expanded, conformal D3-brane solution was performed in \cite{fiol2}. With the non-conformal BPS flow solution, we find that the source term for the dilaton  provided by the D3-brane evaluates to:
\be
J_{\rm D3}\,=\,\frac{N}{2\pi^2}\,\frac{\tilde\kappa}{z^2}\,\sin\psi\,\delta\left(\rho\,-\,\tfrac{\tilde\kappa \,z}{1\,+\,a\tilde\kappa z}\right)\,.
\ee
Here $\psi$ is the polar angle of the spatial two-sphere along the gauge theory directions, wrapped by the D3-brane worldvolume.
Retracing the steps outlined in \cite{fiol2}, but now applied to BPS flow embedding, we find that the leading term in the expansion of the rescaled  dilaton \eqref{canonical} near the boundary $z\to 0$ is:
\bea
\tilde\phi(\rho,\,z)\,\simeq\,\frac{15\sqrt{2}\tilde \kappa}{16\pi}z^4\int_0^\infty d\rho^\prime\int_0^{\pi}d\theta'\,\sin\theta'\int_0^\infty dz^\prime\,z^{\prime \,2}
\frac{\delta\left(\rho'\,-\,\frac{\tilde\kappa\,z'}{1\,+\,a\tilde\kappa\,z'}\right)}{\left(z^{\prime 2}\,+\,(\vec x\,-\vec x^\prime)^2\right)^{\frac{7}{2}}}\,.
\eea
The expectation value of ${\cal O}_{F^2}$ is given by the coefficient of $z^4$. Performing the angular integral, we obtain
\bea
\langle{\cal O}_{ F^2}(\rho)\rangle\,=\,\frac{3\sqrt{2}\tilde \kappa}{8\pi}\int_0^\infty
dz'\,&&\frac{1+a\tilde\kappa z'}{2\rho\tilde\kappa z'}\,z^{\prime\,2}\,\times\\\nonumber\\\nonumber
&&\left[\left(z^{\prime\,2}\,+\,\left(\rho-\tfrac{\tilde\kappa z'}{1+a\tilde\kappa z'}\right)^2\right)^{-\frac{5}{2}}\,-\,\left(z^{\prime\,2}\,+\,\left(\rho+\tfrac{\tilde\kappa z'}{1+a\tilde\kappa z'}\right)^2\right)^{-\frac{5}{2}}\right]\,.
\eea
The integral cannot be evaluated analytically. The limiting cases of $a=0$ and $a=\infty$, corresponding to the symmetric representation and $k$ coincident strings, respectively, are easily computed and yield,
\bea
\langle{{\cal O}_{F^2}}\rangle\,&&=\,\frac{\sqrt 2}{4\pi}\,\frac{\tilde\kappa\sqrt{1+\tilde\kappa^2}}{\rho^4}\,\qquad {\rm for}\quad a\,=\,0\,,\\\nonumber\\\nonumber
&&=\,\frac{\sqrt 2}{4 \pi}\,\frac{\tilde\kappa}{\rho^4}\,\qquad\qquad\quad{\rm for}\quad a\,\to\,\infty\,.
\eea
The two limits respectively control the small $\rho$ and large $\rho$ asymptotics of $\langle {\cal O}_{F^2}\rangle$. We note that our normalisations reproduce the result for $k$ strings obtained from the BPS D5-brane spike \eqref{f^2D5}.  
Figure \ref{d3vev} shows the complete flow for fixed $a$, and confirms that $\langle {\cal O}_{F^2}\rangle$ interpolates between the symmetric representation at short distances and $k$ fundamental quarks in the IR. Both limits yield conformal behaviour $\langle {\cal O}_{F^2}\rangle \sim 1/\rho^4$, whilst the magnitude of the coefficient actually {\em decreases} towards the IR. For this reason, analogously to the D5-brane case, we refer to this as a screening of the source in the symmetric representation.

\subsection{Finite temperature D3-brane embeddings}
\begin{figure}
\begin{centering}
\includegraphics[width=2.5in]{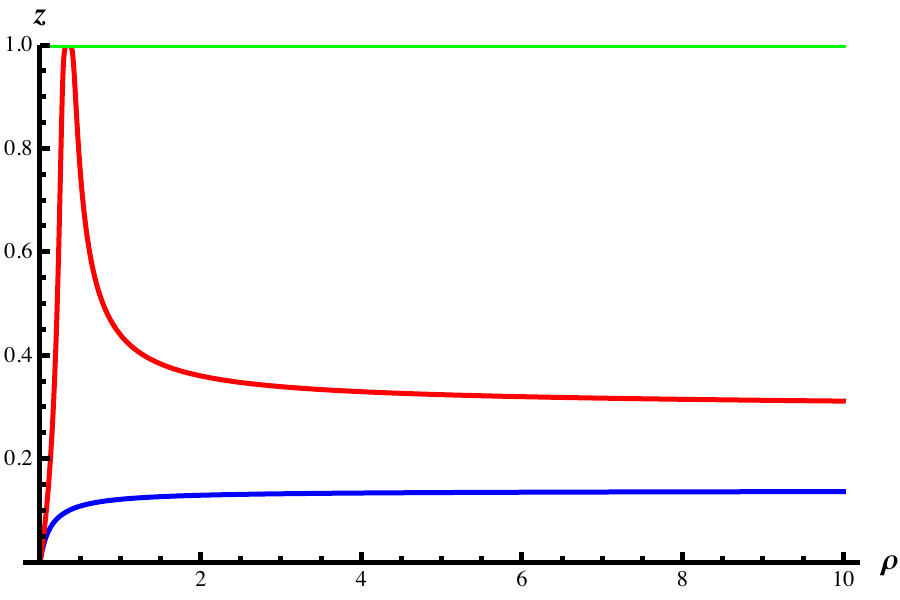}\hspace{0.5in}\includegraphics[width=2.5in]{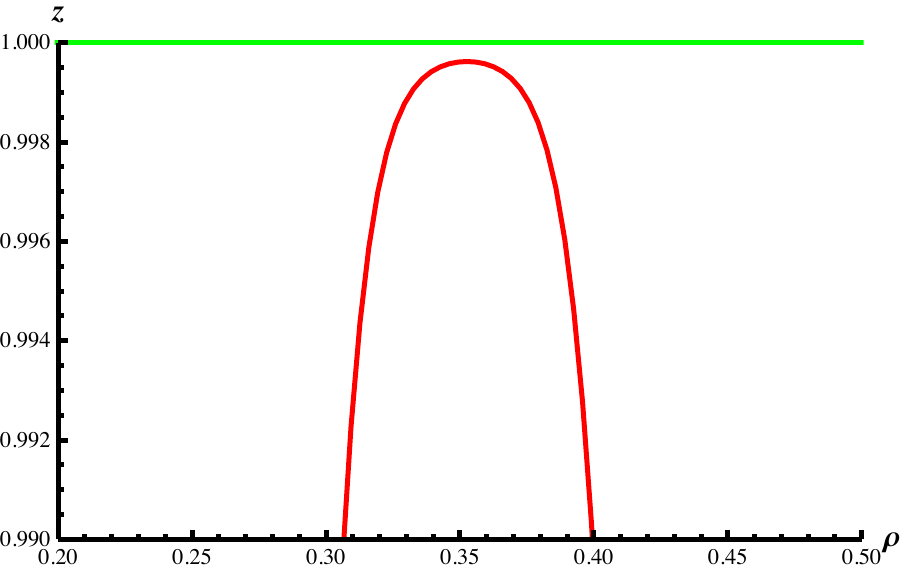}
\caption{\small Solutions to the D3-brane embedding equations at finite temperature for  $\k\,=\,1$ and $z_+\,=\,1/\pi T\,=\,1$. Both the red ($a=4$) and  blue ($a=-8$) curves do not get to the horizon, and instead run off to spatial infinity. Zooming in near $z=z_+=1$ (right) we observe that the red curve never reaches the horizon.}
\label{nonext}
\end{centering}
\end{figure}

Thermal corrections to straight Wilson lines are computed by considering appropriate dual objects embedded in the (Euclidean) AdS-Schwarzschild geometry. Extensive investigations of the D3-brane embedding equations in global AdS$_5$-Schwarzschild geometry (dual to thermal ${\cal N}=4$ theory on the three-sphere) were performed in \cite{us1} and no nontrivial, expanded D3-brane configurations could be found with one end-point on the boundary. The only allowed solution is the collapsed embedding representing $k$ coincident strings.
One may therefore be tempted to argue that any finite temperature (at strong coupling) has the effect of ``dissociating'' the symmetric representation into  fundamental quarks. A similar statement also applies to the D5-branes and the antisymmetric representation above the critical temperature $T_c(\kappa)$, beyond which non-constant D5-brane embeddings appear not  to exist for the gauge theory on the three-sphere.

Below we outline the result of our numerical investigation of non-constant, finite temperature D3-brane embeddings in the {\em planar} AdS-Schwarzschild geometry, describing the thermal CFT on $\mathbb{R}^4$.
For the planar black hole with metric:
\bea
&&ds^2\,=\,\frac{1}{z^2}\left[\fr{dz^2}{\tilde f(z)}\,+\,\tilde f(z)\,dt^2\,+\,d\rho^2\,+\,\rho^2\,d\Omega_{\,2}^{\,2}\right]\,,\\\nonumber\\\nonumber
&&\tilde f(z)\,=\,1\,-\,\left(\fr{z}{z_+}\right)^4\,,\qquad T\,=\,\fr{1}{z_+\,\pi}\,,
\eea
at temperature $T$, the action for the D3-brane embedding is,
\be
S_{\rm D3}\,=\,\fr{2N}{\pi}\int dt\,d\sigma\,\fr{\rho^2}{z^4}\,\left(\sqrt{(\pa_\sigma z)^2\,+\,\tilde f(z)(\pa_\sigma\rho)^2\,-\,G^{\,2}\,z^4}\,-\,\pa_\sigma \rho\right)\,+\,S_{\rm c.t.}\,.
\ee
In this case we work in the gauge $\sigma\,=\,\rho$, and solve the  full second order equations for $z(\rho)$. In terms of the electric  field  $G$ on the brane,
\be
G\,=\,\fr{\tilde \k\sqrt{(\pa_\sigma z)^2\,+\,\tilde f(z)\,(\pa_\sigma \rho)^2}}{\sqrt{\rho^4\,+\,\tilde \k^2\,z^4}}
\ee
 the equation of motion for $z(\rho)$  is,
\be
\fr{2\,\tilde \k\,G}{z}\,+\,\fr{4\,\rho^2}{z^5}\left(\fr{G\,\rho^2}{\tilde\k}\,-\,1\right)\,+\,\fr{d}{d\rho}\left(\fr{\tilde\k\,\pa_{\,\rho}z}{z^4\,G}\right)\,=\,\fr{\partial_\rho\tilde f(z)\,\tilde\k}{2\,z^4\, G}\,.
\ee
We now search for solutions  by expanding about the UV ($z\to 0$)
\be
z(\rho\to 0)\,\simeq\,\fr{\rho}{\tilde\k}\,+\,\frac{a}{\tilde\kappa}\,\rho^2\,+\,\left(\frac{a}{\tilde\kappa}\right)^2\,\tilde\k\,\rho^3\,+\,\left(\frac{a}{\tilde\kappa}\right)^3\,\tilde\k^2\,\,\rho^4\,+\,\fr{6\,\tilde\k^8\,z_+^4\,(a \tilde\kappa^{-1})^4\,-\,1}{6\,z_+^4\,\tilde\k^5}\,\rho^5\,+\ldots\nonumber
\ee
and integrating-in/shooting towards the horizon. There are two distinct behaviours for the solutions, as shown in Figure \ref{nonext},  distinguished by the sign of the deformation parameter: positive $a$ leads to solutions that approach the horizon before turning around and running off to spatial infinity $(\rho\to \infty)$, whereas negative $a$ solutions directly run off to spatial infinity. This qualitative structure appears to be  generic for any non-zero temperature.
We conclude that there are no expanded D3-brane configurations at finite temperature, related to symmetric representation Polyakov loops or flows originating from them.

\section{Discussion}
We have analysed the interpretation and thermodynamics of a class of D3- and D5-brane probe embeddings in AdS$_5\times {\rm S}^5$ which interpolate between Wilson/Polyakov loops in  higher rank tensor representations and the fundamental representation. In both cases we could characterise the nonconformal probe embeddings in terms of a flow induced by a deformation that could be interpreted as a VEV  for a $\Delta=1$ operator in the UV description of the quantum mechanical impurity. The UV limit for the D5-brane embedding corresponds to $k$ fundamental quarks whilst the same limit for the D3-brane is dual to a source  transforming in the  symmetric tensor representation of the gauge group. The D3-brane case is particularly intriguing, since the one-point function of ${\cal O}_{F^2}$ shows a decrease in the strength of the coupling to the source whilst interpolating between a symmetric representation source in the UV and $k$ fundamental quarks in the IR. The ratio $\langle {\cal O}_{F^2}\rangle/\langle {\cal O}_{F^2}\rangle_{\Box}$ decreases monotonically with distance from the source in the boundary gauge theory. Although naturally suggestive of a ``thinning out'' of degrees of freedom towards the IR, other measures e.g. the entanglement entropy (EE) of Wilson line defects \cite{Lewkowycz:2013laa}  and a proposed holographic $g$-function \cite{Yamaguchi:2002pa} do not support this interpretation. In particular, the results of \cite{Lewkowycz:2013laa} for conformal probes imply that the contributions from  symmetric, antisymmetric and fundamental representation  sources to the EE of a spherical domain in the gauge theory (at large $N$ and 't Hooft coupling $\lambda\gg1$) are, respectively:
\be
S^{\rm EE}_\Box\,=\,k\frac{\sqrt\lambda}{3}\,,\qquad S^{\rm EE}_{{\cal A}_k}\,=\frac{2N}{9\pi}\sqrt\lambda\,\sin^3\theta_k\,,\qquad\,S^{\rm EE}_{{\cal S}_k}\,=\,2N\left(\sinh^{-1}\tilde\kappa\,-\,\tfrac{1}{3}\tilde\kappa\sqrt{\tilde\kappa^2+1}\right)\,.\nonumber
\ee
It is easy to check that $S^{\rm EE}_{{\cal A}_k} < S^{\rm EE}_\Box $ {\em and} 
$S^{\rm EE}_{{\cal S}_k} < S^{\rm EE}_\Box$, whilst keeping $\kappa$ fixed in the first case, and $\tilde \kappa$ fixed in the second. For the D3-brane flow, this clashes with the intuitive picture implied by  the behaviour of $\langle{\cal O}_{F^2}\rangle$  (figure \ref{d3vev}) at the fixed points. 
It should be possible to calculate the entanglement entropy for the non-conformal D3- and D5-brane embeddings discussed in this paper using the techniques in \cite{Karch:2014ufa}. The contribution from defects/impurities to the EE of a spherical region of radius $R$ is a candidate $g$-function (see e.g. \cite{Casini:2016fgb, Jensen:2013lxa}) and its behaviour for the D3-brane flow solution would clearly be very interesting. It should also be possible to analyse the D3-brane system from the viewpoint of the bosonic quantum mechanics of the boundary impurity. For D5-branes computing the antisymmetric representation, the corresponding fermionic impurity model has been solved exactly \cite{muck, kondo, Sachdev:2010uj} but a similar analysis for the symmetric representation D3-brane (and deformations) is lacking.

Finally, given the two types of interpolating solutions above, a natural question is whether there exist flows from the symmetric representation (expanded D3-brane) in the UV to the antisymmetric one (expanded D5-brane). A possible approach to this question is via a D7-brane embedding with worldvolume AdS$_2\times{\rm S}^2\times {\rm S}^4$, carrying $k$ units of string charge. 
\acknowledgments
We would like to thank Adi Armoni, Carlos N\'u\~nez and Sanjaye Ramgoolam for stimulating comments and discussions.
The authors were supported in part by STFC grant ST/L000369/1  and ST/K5023761/1.


\newpage
\begin{thebibliography}{99}
\bibitem{maldacena}
  J.~M.~Maldacena,
{  ``The large N limit of superconformal field theories and supergravity,''}
  Adv.\ Theor.\ Math.\ Phys.\  {\bf 2}, 231 (1998)
  [Int.\ J.\ Theor.\ Phys.\  {\bf 38}, 1113 (1999)]
  [hep-th/9711200].
  
\bibitem{gkp} 
  S.~S.~Gubser, I.~R.~Klebanov and A.~M.~Polyakov,
  ``Gauge theory correlators from noncritical string theory,''
  Phys.\ Lett.\ B {\bf 428}, 105 (1998)
  doi:10.1016/S0370-2693(98)00377-3
  [hep-th/9802109].


\bibitem{witten}
  E.~Witten,
{  ``Anti-de Sitter space and holography,''}
  Adv.\ Theor.\ Math.\ Phys.\  {\bf 2}, 253 (1998)
  [hep-th/9802150].\\
  
  
 
\bibitem{malwil} 
  J.~M.~Maldacena,
  ``Wilson loops in large N field theories,''
  Phys.\ Rev.\ Lett.\  {\bf 80}, 4859 (1998)
  doi:10.1103/PhysRevLett.80.4859
  [hep-th/9803002]; 
  S.~J.~Rey and J.~T.~Yee,
  ``Macroscopic strings as heavy quarks in large N gauge theory and anti-de Sitter supergravity,''
  Eur.\ Phys.\ J.\ C {\bf 22}, 379 (2001)
  doi:10.1007/s100520100799
  [hep-th/9803001].

\bibitem{Rey:1998bq} 
  S.~J.~Rey, S.~Theisen and J.~T.~Yee,
  ``Wilson-Polyakov loop at finite temperature in large N gauge theory and anti-de Sitter supergravity,''
  Nucl.\ Phys.\ B {\bf 527}, 171 (1998)
  doi:10.1016/S0550-3213(98)00471-4
  [hep-th/9803135].
  
  
\bibitem{Witten:1998zw}
  E.~Witten,
  ``Anti-de Sitter space, thermal phase transition, and confinement in gauge theories,''
  Adv.\ Theor.\ Math.\ Phys.\  {\bf 2} (1998) 505
  [hep-th/9803131].



\bibitem{fiol}
  N.~Drukker and B.~Fiol,
  ``All-genus calculation of Wilson loops using D-branes,''
  JHEP {\bf 0502} (2005) 010
  [hep-th/0501109].



\bibitem{us1}
  S.~A.~Hartnoll and S.~P.~Kumar,
  ``Multiply wound Polyakov loops at strong coupling,''
  Phys.\ Rev.\ D {\bf 74} (2006) 026001
  [hep-th/0603190].


\bibitem{yamaguchi}
  S.~Yamaguchi,
  ``Wilson loops of anti-symmetric representation and D5-branes,''
  JHEP {\bf 0605} (2006) 037
  [hep-th/0603208].


\bibitem{Hartnoll:2006is} 
  S.~A.~Hartnoll and S.~P.~Kumar,
  ``Higher rank Wilson loops from a matrix model,''
  JHEP {\bf 0608}, 026 (2006)
  doi:10.1088/1126-6708/2006/08/026
  [hep-th/0605027].


\bibitem{passerini}
  J.~Gomis and F.~Passerini,
  ``Holographic Wilson Loops,''
  JHEP {\bf 0608} (2006) 074
  [hep-th/0604007].


  
\bibitem{muck} 
  W.~Mueck,
  ``The Polyakov Loop of Anti-symmetric Representations as a Quantum Impurity Model,''
  Phys.\ Rev.\ D {\bf 83}, 066006 (2011)
  Erratum: [Phys.\ Rev.\ D {\bf 84}, 129903 (2011)]
  doi:10.1103/PhysRevD.83.066006, 10.1103/PhysRevD.84.129903
  [arXiv:1012.1973 [hep-th]].


\bibitem{kondo} 
  S.~Harrison, S.~Kachru and G.~Torroba,
  ``A maximally supersymmetric Kondo model,''
  Class.\ Quant.\ Grav.\  {\bf 29}, 194005 (2012)
  doi:10.1088/0264-9381/29/19/194005
  [arXiv:1110.5325 [hep-th]].
  
\bibitem{Sachdev:2010uj} 
  S.~Sachdev,
  ``Strange metals and the AdS/CFT correspondence,''
  J.\ Stat.\ Mech.\  {\bf 1011}, P11022 (2010)
  doi:10.1088/1742-5468/2010/11/P11022
  [arXiv:1010.0682 [cond-mat.str-el]].

\bibitem{Callan:1998iq}
  C.~G.~Callan, Jr., A.~Guijosa and K.~G.~Savvidy,
  { ``Baryons and string creation from the five-brane world volume action,''}
  Nucl.\ Phys.\ B {\bf 547}, 127 (1999)
  [hep-th/9810092].

\bibitem{Camino:2001at} 
  J.~M.~Camino, A.~Paredes and A.~V.~Ramallo,
  ``Stable wrapped branes,''
  JHEP {\bf 0105}, 011 (2001)
  doi:10.1088/1126-6708/2001/05/011
  [hep-th/0104082].

\bibitem{Drukker:1999zq} 
  N.~Drukker, D.~J.~Gross and H.~Ooguri,
  ``Wilson loops and minimal surfaces,''
  Phys.\ Rev.\ D {\bf 60}, 125006 (1999)
  doi:10.1103/PhysRevD.60.125006
  [hep-th/9904191].
  
  
\bibitem{Klebanov:1997kc} 
  I.~R.~Klebanov,
  ``World volume approach to absorption by nondilatonic branes,''
  Nucl.\ Phys.\ B {\bf 496}, 231 (1997)
  doi:10.1016/S0550-3213(97)00235-6
  [hep-th/9702076].

\bibitem{Callan:1999ki}
  C.~G.~Callan, Jr. and A.~Guijosa,
  {``Undulating strings and gauge theory waves,''}
  Nucl.\ Phys.\ B {\bf 565}, 157 (2000)
  [hep-th/9906153]; 



\bibitem{Danielsson:1998wt}
  U.~H.~Danielsson, E.~Keski-Vakkuri and M.~Kruczenski,
  { ``Vacua, propagators, and holographic probes in AdS / CFT,''}
  JHEP {\bf 9901}, 002 (1999)
  [hep-th/9812007].

\bibitem{fiol2} 
  B.~Fiol, B.~Garolera and A.~Lewkowycz,
  ``Exact results for static and radiative fields of a quark in N=4 super Yang-Mills,''
  JHEP {\bf 1205}, 093 (2012)
  doi:10.1007/JHEP05(2012)093
  [arXiv:1202.5292 [hep-th]].


\bibitem{Imamura:1998gk}
  Y.~Imamura,
  ``Supersymmetries and BPS configurations on Anti-de Sitter space,''
  Nucl.\ Phys.\ B {\bf 537} (1999) 184
  [hep-th/9807179].


\bibitem{schwarz} 
  J.~H.~Schwarz,
  ``BPS Soliton Solutions of a D3-brane Action,''
  JHEP {\bf 1407}, 136 (2014)
  doi:10.1007/JHEP07(2014)136
  [arXiv:1405.7444 [hep-th]].

\bibitem{Faraggi:2011bb}
  A.~Faraggi and L.~A.~Pando Zayas,
  ``The Spectrum of Excitations of Holographic Wilson Loops,''
  JHEP {\bf 1105} (2011) 018
  [arXiv:1101.5145 [hep-th]].

\bibitem{Faraggi:2011ge}
  A.~Faraggi, W.~Mueck and L.~A.~Pando Zayas,
  ``One-loop Effective Action of the Holographic Antisymmetric Wilson Loop,''
  Phys.\ Rev.\ D {\bf 85} (2012) 106015
  [arXiv:1112.5028 [hep-th]].
  
\bibitem{Faraggi:2014tna} 
  A.~Faraggi, J.~T.~Liu, L.~A.~Pando Zayas and G.~Zhang,
  ``One-loop structure of higher rank Wilson loops in AdS/CFT,''
  Phys.\ Lett.\ B {\bf 740}, 218 (2015)
  doi:10.1016/j.physletb.2014.11.060
  [arXiv:1409.3187 [hep-th]].

\bibitem{Lewkowycz:2013laa} 
  A.~Lewkowycz and J.~Maldacena,
  ``Exact results for the entanglement entropy and the energy radiated by a quark,''
  JHEP {\bf 1405}, 025 (2014)
  doi:10.1007/JHEP05(2014)025
  [arXiv:1312.5682 [hep-th]].
  
\bibitem{Yamaguchi:2002pa}
  S.~Yamaguchi,
  ``Holographic RG flow on the defect and g theorem,''
  JHEP {\bf 0210} (2002) 002
  doi:10.1088/1126-6708/2002/10/002
  [hep-th/0207171].
  

  
\bibitem{Karch:2014ufa} 
  A.~Karch and C.~F.~Uhlemann,
  ``Generalized gravitational entropy of probe branes: flavor entanglement holographically,''
  JHEP {\bf 1405}, 017 (2014)
  doi:10.1007/JHEP05(2014)017
  [arXiv:1402.4497 [hep-th]].
  
  
\bibitem{Casini:2016fgb} 
  H.~Casini, I.~S.~Landea and G.~Torroba,
  ``The g-theorem and quantum information theory,''
  JHEP {\bf 1610}, 140 (2016)
  doi:10.1007/JHEP10(2016)140
  [arXiv:1607.00390 [hep-th]].
  
\bibitem{Jensen:2013lxa} 
  K.~Jensen and A.~O'Bannon,
  ``Holography, Entanglement Entropy, and Conformal Field Theories with Boundaries or Defects,''
  Phys.\ Rev.\ D {\bf 88}, no. 10, 106006 (2013)
  doi:10.1103/PhysRevD.88.106006
  [arXiv:1309.4523 [hep-th]].
  
 \end{thebibliography}
\end{document}